# Quantum dynamic and geometric phases in harmonic oscillator, spin 1/2 and two-level thermal engines systems

Y. Ben-Aryeh

*Technion-Israel Institute of Technology, Physic Department, Israel, Haifa, 32000*

*Email: phr65yb@physics.technion.ac.il*

**Abstract**

Dynamical phases are obtained for a quantum thermal engine, whose working medium is a single harmonic oscillator. The dynamics of this engine is obtained by using four steps where in two steps the time dependent frequency is changing, from $\omega_1(\omega_2)$ to $\omega_2(\omega_1)$. In the other two steps, the thermal engine is coupled alternatively to hot and cold heat baths with temperatures $T_H$ and $T_C$, respectively. Similar dynamical phases are obtained in a quantum thermal engine whose working medium is spin 1/2 system. The role of times durations of such steps in the quantum engines for getting maximal efficiency is analyzed. The dynamic of charge pumping in a quantum dot coupled to two reservoirs is studied. The effects of many modulation parameters including their fluctuations for getting geometric phases are analyzed. Since the separate steps in thermal engines describe non-cyclic circuits, we propose to use a special method for measuring geometric phases in thermal engines for non-cyclic circuits which is gauge invariant.

## 1. Introduction

Quantum thermal engines connected to a hot and cold heat bath have been studied extensively in various forms. The most studied such engines are based on Otto [1-4] and Carnot [5-8] cycles. The relations between these engines and thermodynamics and their efficiency were investigated. A single harmonic oscillator as working medium for such engines is analytically solvable model. It has therefore been studied extensively and motivated experimental realization [3]. This device can be classified as Otto engine operating by periodically modulating the trap frequency.

While the quantum properties of time dependent Otto cycles were studied in many works the emphasis in those studies was on the efficiency of such engines. We raise in the present work the question: if and how geometric and dynamical phases are involved in such cycles. This question is of much importance as by using interference effects we might obtain important conclusions on the quantum or classical nature of such systems.

We distinguish between dynamical and geometrical phases by following a simple theory [9,10]: If a quantum system is initially in an eigenstate of the Hamiltonian $H$ and is changed with time, according to the slowly varying parameters, the adiabatic conditions guarantee that it remains an eigenstate of the instantaneous Hamiltonian. Assuming an initially eigenstate of the Hamiltonian

$$H(\vec{R})|n,\vec{R}\rangle = E_n(\vec{R})|n,\vec{R}\rangle \tag{1}$$

and parameters $\vec{R}$ which are slowly varying along a closed curve $C$ in parameter space at time $T$, the geometrical phase factor is given by [9,10]

$$\gamma_n(C) = i\oint d\vec{R} \cdot \langle n,\vec{R}|\nabla_{\vec{R}}|n,\vec{R}\rangle \tag{2}$$

This phase is given in addition to the dynamical phase given by

$$\phi_{dyn} = -\int_0^T E_n(t)dt \quad . \tag{3}$$

We find in the present analysis for harmonic oscillator thermal engines that the quantum phases are related to dynamical and entropic phases. Also, in the present analysis for Carnot cycle of spin 1/2 systems quantum phases are 'dynamical' without geometrical properties.

It has been shown, however, by extensive analysis [11] that many observables characterizing the performance of the thermal machines are of geometric nature. Especially it has been shown that the antisymmetric component of the thermal tensor can be identified as a Berry curvature. While such results are of much interest, in many works on the *efficiency* of thermal engines only dynamical phases are relevant (as will be illustrated in the analysis of Sections 2 and 3). This fact is explained by the performance of thermal engines decomposed into steps where each step includes only one modulating parameter. But it is possible by adiabatic driving to pump heat in thermal machines by many slowly modulating parameters periodically in time. Such modulating parameters can lead to geometrical phases in addition to the dynamical phases.

In Section 4 we describe thermal engines which include many modulating parameters related to stochastic systems describing the interaction of a dot with two reservoirs and leading to both dynamical and geometric phases. In section 5 we discuss the possibility to get geometric phases in spin 1/2 thermal engine system with 'non-cyclic' circuits. In Section 6, we summarize our results and conclusions.

## 2. Reversible and irreversible harmonic oscillator Otto cycle

We consider first a quantum thermal engine whose working medium is a single harmonic oscillator with time dependent frequency changing between $\omega_1$ and $\omega_2$ [4]. The engine is alternatively coupled to two heat baths at hot temperatures $T_H$ and cold temperature $T_C$.

There is a trade between maximum power, and maximum efficiency at which power is zero. Therefore, the harmonic oscillator Otto cycle and correspondingly its quantum phases can be analyzed under different limiting cases: In case A we treat the limiting case of Otto cycle which works under reversible conditions. In this case the power is tending to zero. In case B we treat the irreversible Otto cycle, which gives maximal power but with energy losses.

### Case A: Reversible Otto cycle

When the parameters are changed *infinitely* slowly along some path from a certain initial point to a final point in parameter space, then the total work $W$ performed on the system is equal to the Helmholtz free energy difference $\Delta F$ [12]. The Helmholtz free energy is defined as $F = U - TS$ where $U$ is the internal energy, T is the absolute temperature (Kelvins) and $S$ is the entropy. We follow here the analysis given in [4] but we relate the quantum phases to those which are dynamical.

The Otto cycle consists of four consecutive steps (see Figure 1 of [4]).

1. In the first step of isentropic compression the frequency of the oscillator is changed between points A and B from $\omega_1$ to $\omega_2$ but its temperature $T_H$ is not changed as the system is isolated and work is performed only by modulating the frequency. The mean work done during the first step between these points is given by

$$\langle W_1 \rangle = \langle H \rangle_B - \langle H \rangle_A = \left( \frac{\hbar\omega_2}{2} - \frac{\hbar\omega_1}{2} \right) \coth\left( \frac{\beta_1 \hbar\omega_1}{2} \right) \quad . \tag{4}$$

where $\beta_1 = \dfrac{1}{k_B T_H}$ and $k_B$ is the Boltzmann constant.

2. In the second step of hot isochore heat is transferred from the hot bath to the oscillator. The temperature of the oscillator is changing between points B and C from $T_H$ to $T_C$ but its frequency $\omega_2$ is not changing. The mean heat $\langle Q_2 \rangle$ exchanged with the hot bath in the second step is given by

$$\langle Q_2 \rangle = \langle H \rangle_C - \langle H \rangle_B = \frac{\hbar\omega_2}{2}\left[ \coth\left( \frac{\beta_2 \hbar\omega_2}{2} \right) - \coth\left( \frac{\beta_1 \hbar\omega_1}{2} \right) \right] \tag{5}$$

where $\beta_2 = \dfrac{1}{k_B T_C}$

In the third step of isentropic expansion the frequency of the oscillator is changed between points C and D from $\omega_2$ to $\omega_1$ but the temperature $T_C$ is not changed.. The average work done in the third step is given by

$$\langle W_3 \rangle = \langle H \rangle_D - \langle H \rangle_C = \left( \frac{\hbar\omega_1}{2} - \frac{\hbar\omega_2}{2} \right) \coth\left( \frac{\beta_2 \hbar\omega_2}{2} \right) \quad . \tag{6}$$

4. In the fourth step of cold isochore heat is transferred from the oscillator to the cold bath. At this step the temperature of the oscillator is changed between points D and A from $T_H$ to

$T_C$ but its frequency $\omega_1$ is not changed. The average heat transferred to the cold bath is given by

$$\langle Q_4 \rangle = \langle H \rangle_A - \langle H \rangle_D = \frac{\hbar \omega_1}{2} \left[ \coth\left(\frac{\beta_1 \hbar \omega_1}{2}\right) - \coth\left(\frac{\beta_2 \hbar \omega_2}{2}\right) \right]. \tag{7}$$

By dividing $-\left[\langle W_1 \rangle + \langle W_3 \rangle\right]$ (representing the work obtained in the Otto circuit) by $Q_2$ given by Eq. (5) we get the efficiency of this Otto circle as

$$\eta = -\frac{\langle W_1 \rangle + \langle W_3 \rangle}{Q_2} = 1 - \frac{\omega_1}{\omega_2}. \tag{8}$$

### Case B for irreversible Otto cycle with maximal power

We consider finite-time cycles which moves the engine away from the reversible regime. We define $\tau_C$ ($\tau_H$) the time duration which the system is in contact with the cold (hot) bath along a cycle. There are various proximations which take in account the finite time duration of the process. Such approximations for treating the above irreversible Otto cycle were described in the literature. Here we explain a certain important approximation.

We consider the high-temperature regime for which $\beta_j \hbar \omega_j \ll 1$ ($j = 1, 2$). Under this condition Eqs. (4) and (6) obtain the simple form:

$$\langle W_1 \rangle = \left(\frac{\hbar \omega_2}{2} - \frac{\hbar \omega_1}{2}\right) \frac{2}{\beta_1 \hbar \omega_1} \quad ; \quad \langle W_3 \rangle = \left(\frac{\hbar \omega_1}{2} - \frac{\hbar \omega_2}{2}\right) \frac{2}{\beta_2 \hbar \omega_2} . \tag{9}$$

Then we get

$$2\frac{\langle W_1 \rangle + \langle W_3 \rangle}{\hbar} = \frac{1}{\beta_1}\left(\frac{\omega_2}{\omega_1} - 1\right) + \frac{1}{\beta_2}\left(\frac{\omega_1}{\omega_2} - 1\right) . \tag{10}$$

Assuming fixed values for $\beta_1$ and $\beta_2$, then the maximal value of $\langle W_1 \rangle + \langle W_3 \rangle$ is obtained for:

$$\frac{\omega_2}{\omega_1} = \sqrt{\frac{\beta_1}{\beta_2}} \qquad (11)$$

By substituting Eq. (11) into Eq. (8) we get for the efficiency at maximum power:

$$\eta = 1 - \sqrt{\frac{\beta_2}{\beta_1}} = 1 - \sqrt{\frac{T_C}{T_H}} \qquad (12)$$

which corresponds to Curzon-Ahlborn-efficiency [8]. One can develop other approximations but our main interest in the present work is in the dynamical and geometric phases.

## Quantum phases in harmonic oscillator Otto cycle

In the present reversible Otto cycle, there are not any geometrical phase changes. By comparing Eq. (3) with Eqs. (4) and (6) (given by $\langle W_1 \rangle$ and $\langle W_3 \rangle$)) we find that the latter equations represent dynamical phase changes. We find also in the irreversible otto cycle that these phases decay inversely proportional to time durations of these steps (For duration time tending to infinity we recover the reversible condition). In the reversible Otto cycle Eqs. (5) and (7) represent entropy phase changes, similar to dynamical (without any geometric phase changes). Under the irreversible conditions in the high temperature limit the special value of these phases along the Otto circle are given by Eq. (12). In more general treatment of thermal engines, one should include also different kinds of energy losses, but we included above only the most important factors.

## 3. Reversible and irreversible Carnot cycle with spin 1/2 system

We consider here a quantum heat engine between two heat reservoirs, at constant hot temperature $T_H$ and cold temperature $T_C$, where the spin-1/2 system is coupled with the reservoirs including a magnetic field which can be changed over time [6,7]. The magnetic field $\vec{B}$ is along the positive $z$ axis and the Hamiltonian are given by

$$\hat{H}(t) = 2\mu_B B_z(t)\hat{s}_z \equiv \tilde{\omega}(t)\hat{s}_z \quad ; \quad \tilde{\omega}(t) = 2\mu_B B_z(t) \qquad (13)$$

where here the units are such that $\hbar = 1$, $\mu_B$ is the Bohr magneton and $\tilde{\omega}(t)$ is proportional to $B_z(t)$. We refer to $\tilde{\omega}(t)$ rather than $B_z(t)$ as "the field". Based on statistical mechanics, the expectation value of $\hat{S}_z$ is given by

$$\langle s_z \rangle = -\frac{1}{2}\tanh(\tilde{\omega}\beta/2) \quad ; \quad \beta = \frac{1}{k_b T} \tag{14}$$

Internal energy $E$ of the spin system is given by the expectation value of the Hamiltonian:

$$E = \langle H \rangle = \tilde{\omega}\langle s_z \rangle \tag{15}$$

We describe first the reversible cycle of the Carnot type [5].

## Case A: Reversible Carnot cycle

The present spin cycle (see Figure 1 in [6]) consists of four consecutive steps where two steps are isotherms and the other two are adiabats. The reversible hot isotherm in the $(\tilde{\omega}, \langle s_z \rangle)$ plane satisfies the equation

$$\langle s_z \rangle_H = -\frac{1}{2}\tanh(\tilde{\omega}\beta_H/2) \quad ; \quad \beta_H = \frac{1}{k_B T_H} \tag{16}$$

The first step starts at point A with polarization $\langle s_z \rangle_1$, field $\tilde{\omega}_1$ and ending at point B with polarization $\langle s_z \rangle_2$ and field $\tilde{\omega}_2$. The absorbed heat $Q_H$ by the spin system (including $\hbar$) from the hot reservoir is given by

$$Q_H = \langle H_B \rangle - \langle H_A \rangle = \frac{\hbar}{2}\left[\tilde{\omega}_2 \tanh(\beta_H \hbar \tilde{\omega}_2) - \tilde{\omega}_1 \tanh(\beta_H \hbar \tilde{\omega}_1)\right] \tag{17}$$

where $\langle H_B \rangle$ and $\langle H_A \rangle$ are the expectation values for the Hamiltonian at points B and A, respectively, given by Eq. (13) for the hot isotherm.

In the second step the work $\langle W_1 \rangle$ was obtained as

$$\langle W_1 \rangle = \langle H_C \rangle - \langle H_B \rangle = \frac{\hbar}{2} \left[ \tilde{\omega}_3 \tanh(\beta_C \hbar \tilde{\omega}_3) - \tilde{\omega}_2 \tanh(\beta_H \hbar \tilde{\omega}_2) \right] \quad . \tag{18}$$

The reversible cold isotherm in the $(\tilde{\omega}, \langle s_z \rangle)$ plane satisfies the equation

$$\langle s_z \rangle_C = -\frac{1}{2} \tanh(\hbar \tilde{\omega} \beta_C / 2) \quad ; \quad \beta_C = \frac{1}{k_B T_C} \quad . \tag{19}$$

The third step starts at point C with polarization $\langle s_z \rangle_2$, field $\tilde{\omega}_3$ and ending at point D with polarization $\langle s_z \rangle_1$ and field $\tilde{\omega}_4$. The emitted heat $Q_C$ from the spin system to the cold reservoir is given by

$$Q_C = \langle H_D \rangle - \langle H_C \rangle = \frac{\hbar}{2} \left[ \tilde{\omega}_4 \tanh(\beta_C \hbar \omega_4) - \tilde{\omega}_3 \tanh(\beta_C \hbar \omega_3) \right] \tag{20}$$

where $\langle H_D \rangle$ and $\langle H_C \rangle$ are the expectation values for the Hamiltonian at points B and A, respectively, given by Eq. (13) and the conditions for the cold isotherm.

In the fourth step the work $\langle W_3 \rangle$ was obtained as

$$\langle W_3 \rangle = \langle H_A \rangle - \langle H_D \rangle = \frac{\hbar}{2} \left[ \tilde{\omega}_1 \tanh(\beta_H \hbar \tilde{\omega}_1) - \tilde{\omega}_4 \tanh(\beta_C \hbar \omega_4) \right] \quad . \tag{21}$$

The efficiency of the Carnot cycle is calculated as

$$\eta = \frac{Q_C + Q_H}{Q_C} \quad ; \quad Q_H = \frac{\hbar}{2} \left[ \tilde{\omega}_2 \tanh(\beta_H \hbar \tilde{\omega}_2) - \tilde{\omega}_1 \tanh(\beta_H \hbar \tilde{\omega}_1) \right]$$
$$Q_C = \frac{\hbar}{2} \left[ \tilde{\omega}_4 \tanh(\beta_C \hbar \omega_4) - \tilde{\omega}_3 \tanh(\beta_C \hbar \omega_3) \right] \tag{22}$$

At points B and C, we have the same $\langle s_z \rangle_2$ value and at points A and D we have the same $\langle s_z \rangle_1$. Then we get after straightforward calculations:

$$\beta_H \tilde{\omega}_1 = \beta_C \tilde{\omega}_4 \quad ; \quad \beta_H \tilde{\omega}_2 = \beta_C \tilde{\omega}_3 \quad ; \quad \frac{\tilde{\omega}_4}{\tilde{\omega}_1} = \frac{\tilde{\omega}_3}{\tilde{\omega}_2} = \frac{\beta_H}{\beta_C} = \frac{T_C}{T_H} \quad . \tag{23}$$

Substituting Eq. (23) in Eq. (22) we get the result for reversible Carnot cycle

$$\eta_{rev} = 1 - \frac{T_C}{T_H} \quad . \tag{24}$$

## Case B: Irreversible Carnot cycle

There are various works treating the irreversibility of the Carnot cycle. We refer here to [5] where a simple model (without entering in the detailed interactions) was developed which gives fair comparisons with experimental results under different conditions. We use here the idea that the irreversible heat from the cold (hot) reservoir will decay with $\tau_C (\tau_H)$ as

$$Q_{ir,C} = T_C \left( -\Delta S - \frac{C_1}{\tau_C} \right); \quad Q_{ir,H} = Q_{r,H} \left( \Delta S - \frac{C_2}{\tau_H} \right) \quad . \tag{25}$$

Here $S$ represents the entropy, $C_1$ and $C_2$ are certain constants and the reversible regime is approached in times $\tau_C \to \infty$ and $\tau_H \to \infty$. We consider the power generated during the Carnot cycle and by using Eq. (25) we get:

$$P = \frac{-W}{\tau_H + \tau_C} = \frac{(T_H - T_C)\Delta S - T_C (C_1/\tau_C) - T_H (C_2/\tau_H)}{\tau_H + \tau_C} \quad . \tag{26}$$

The maximum power is found by setting the derivatives of $P$ with respect to $\tau_H$ and $\tau_C$ equal to zeros. Such procedure was developed in [5] obtaining different results including the case $C_1 = C_2$ for which Curzon-Ahlborn efficiency [8] was obtained.

Eqs. (17-21) describe reversible entropic phase changes while Eqs. (25) and (26) describe irreversibility, i. e. decay of the amplitude, of these phases. All these phases have dynamical nature without geometric properties. In the next Sections we describe certain phases of two-level thermal engines which have geometric properties.

# 4. Geometric phases in stochastic chemical kinetics of a quantum dot coupled to two reservoirs

We consider the dynamic of charge pumping in quantum dot coupled to two reservoirs in the left ($L$) and right ($R$) sides of the dot, described by master equation of the form

$$(d/dt)P(t) = L(t)P(t) \quad ; \quad |P(t)\rangle = (P_0(t), P_1(t))^T \qquad . \tag{27}$$

Here $P_0(t), P_1(t)$ are the probabilities of having 0 or 1 electron on the dot. This system is subjected to a periodic external drive described by the periodic rate matrix:

$$L(t) = l(t+T) \quad ; \quad T = 2\pi/\Omega \tag{28}$$

where $T$ is the period of the drive.

The rate matrix can be given by [13-15]:

$$L(t) = \begin{pmatrix} -\Gamma_L^+(t) - \Gamma_R^+(t) & \Gamma_L^-(t) + \Gamma_R^-(t) \\ \Gamma_L^+(t) + \Gamma_R^+(t) & -\Gamma_L^-(t) - \Gamma_R^-(t) \end{pmatrix} \qquad . \tag{29}$$

Here $\Gamma_\alpha^\pm$ are the time-dependent rates for an electron to tunnel on (+) or off (-) the quantum dot via the left or right reservoir, $\alpha = L, R$. The pumped charge per period into the right reservoir is given by

$$\langle n_R \rangle = \int_0^T dt \left[ \Gamma_R^-(t) P_1(t) - \Gamma_R^+(t) P_0(t) \right] \tag{30}$$

and into the left reservoir by:

$$\langle n_L \rangle = \int_0^T dt \left[ \Gamma_L^-(t) P_1(t) - \Gamma_L^+(t) P_0(t) \right] \qquad . \tag{31}$$

We define the matrices

$$J_+ = \begin{pmatrix} 0 & \Gamma_R^- + \Gamma_L^- \\ 0 & 0 \end{pmatrix} \quad ; \quad J_- = \begin{pmatrix} 0 & 0 \\ \Gamma_L^+ + \Gamma_R^+ & 0 \end{pmatrix} \qquad . \tag{32}$$

Then the pumped charge can be written as

$$\langle n \rangle = \langle n_R \rangle + \langle n_L \rangle = \langle I | J(t) | P(t) \rangle \; ; \; J(t) = J_+(t) - J_-(t) \; ; \; |I\rangle = \begin{pmatrix} 1 & 0 \\ 0 & 1 \end{pmatrix} \quad . \tag{33}$$

We consider adiabatic charge pumping for slow driving with

$$|P(t)\rangle = |\pi(t)\rangle + |\delta\pi(t)\rangle \quad . \tag{34}$$

By substituting the stationary value $|\pi(t)\rangle$ of Eq. (34) into Eq. (33) and integrating over time we get the dynamic pumped charge:

$$N_{dyn} = \int_0^T dt \, \langle I | J(t) | \pi(t) \rangle \quad . \tag{35}$$

We assume that the stationary state of the quantum dot $|\pi(t)\rangle$ satisfies the equation

$$L(t) |\pi(t)\rangle = 0 \quad . \tag{36}$$

The fluctuating terms are related to the master equation

$$\left[ L(t) - (d/dt) \right] \left[ |\pi(t)\rangle + |\delta\pi(t)\rangle \right] = 0 \tag{37}$$

where we are treating the time-derivative as a perturbation of the instantaneous stationary value defined by Eq. (36). We evaluate the periodic sate perturbatively in first order obtaining

$$(d/dt)|\pi(t)\rangle = L(t)|\delta\pi(t)\rangle \rightarrow |\delta\pi(t)\rangle = R(t)(d/dt)|\pi(t)\rangle \; ; \; R(t) = L^{-1}(t) \tag{38}$$

where R(t) is the inverse of L(t). Substituting the fluctuating term $|\delta\pi(t)\rangle$ from Eq. (38) into Eq. (33) we get the geometric part

$$N_{geom} = \int_0^T dt \, \langle I | J(t) R(t) | \partial_t \pi(t) \rangle \quad . \tag{39}$$

To obtain more explicit form of Eq. (39) we find that the time dependence of $J(t), R(t)$ and $\pi(t)$ enters implicitly through the transition rates, $\Gamma = \left(\Gamma_R^+, \Gamma_L^+, \Gamma_R^-, \Gamma_L^-\right)$. Therefore, we can write the integral of Eq. (39) over period as an integral along the closed contour $C$ in the parameter space

$$N_{geom} = \int_0^T d\Gamma \cdot \langle I | J(\Gamma(t)) R(\Gamma(t)) (\partial/\partial \Gamma) | \pi(\Gamma(t)) \rangle = \int_0^T d\Gamma \cdot A(\Gamma) \qquad (40)$$

where $A(\Gamma) = J(\Gamma) R(\Gamma) (\partial/\partial \Gamma) \pi(\Gamma)$ is the vector potential. The above derivations were limited to adiabatic driving. More general derivations including explicit expressions for the $\Gamma$ parameters and their effects were developed in the literature (see e.g. [13-15]}.

## 5. Dynamical and geometric phases in non-cyclic circuits of spin 1/2 thermal engines

Carnot spin 1/2 thermal engines were described in Section 3 by four consecutive steps where two steps are isotherms and the other two are adiabats. Since only dynamic phases were calculated such calculations could be done separately for each step. In the present Section we raise the question if the separate steps of spin 1/2 thermal engine can include also geometric phases. The crucial point is that usually geometric phases are defined in closed contours in order that they will be gauge invariant. But we can extend the definition of geometric phases also for non-cyclic evolution by using the following analysis [10,16,17]: Let us assume that under the adiabatic approximation the initial state $|n, \vec{R}(0)\rangle$ at time $t = 0$ develop after time $t$ into the state $|n, \vec{R}(t)\rangle$ where this development is not cyclic, i.e., $\vec{R}(t) \neq \vec{R}(0)$. We define the geometric phase as

$$\gamma_n(C) = \arg \langle n, \vec{R}(0) | n, \vec{R}(t) \rangle + i \int_{\vec{R}(0)}^{\vec{R}(t)} d\vec{R} \cdot \langle n, \vec{R} | \vec{\nabla}_{\vec{R}} | n, \vec{R} \rangle \qquad (41)$$

where now $C$ is not a closed circuit. We have the following relations

$$\operatorname{Re}\langle n, \vec{R} | \vec{\nabla}_{\vec{R}} | n, \vec{R} \rangle = 0$$
$$\langle n, \vec{R} | \vec{\nabla}_{\vec{R}} | n, \vec{R} \rangle = i \operatorname{Im}\langle n, \vec{R} | \vec{\nabla}_{\vec{R}} | n, \vec{R} \rangle \quad . \tag{42}$$

Therefore, we can write Eq. (41) as

$$\gamma_n(C) = \arg\langle n, \vec{R}(0) | n, \vec{R}(t) \rangle - \operatorname{Im} \int_{\vec{R}(0)}^{\vec{R}(t)} d\vec{R} \cdot \langle n, \vec{R} | \vec{\nabla}_{\vec{R}} | n, \vec{R} \rangle \quad . \tag{43}$$

The crucial point here that $\gamma_n(C)$ is gauge invariant since the gauge changes in the two terms in the right-hand side of equation (43) are equal in magnitude and opposite in sign.

Following the analysis of Section 3 we find that any spin 1/2 state can be represented by the two coordinates: $\tilde{\omega}$ and $\langle s_z \rangle = -\frac{1}{2}\tanh(\tilde{\omega}\beta/2)$ (we find that $\beta$ is a function of these two coordinates). In the description of the four steps of Carnot heat engine in Section 3 we assumed that in each step, one of the two coordinates was constant so that only dynamical phases were obtained. In more general description we can describe steps of Carnot heat engine where the two above coordinates are changed at the same time along the quantum route. For such steps we can use Eq. (43) assuming two-dimensional diagram with coordinates

$$R_1 = \tilde{\omega} \quad ; \quad R_2 = \langle s_z \rangle \tag{44}$$

where each state is described by a certain point in this diagram. The initial state and the final state in such diagram can be described, respectively, as

$$\vec{R}(0) = (\tilde{\omega}, \langle s_z \rangle)_{t=0} \quad ; \quad \vec{R}(t) = (\tilde{\omega}, \langle s_z \rangle)_t \tag{45}$$

The gradient in Eqs. (41,43) can include fluctuations (e. g. like those described in Section 4). For getting optimal efficiency the steps used in Section 3 seem to be optimal but the possibility to get geometric phases by choosing different routes should be of interest. Also It is interesting to note, that a way to measure such phases was described in [17].

14## 6. Summary and conclusions

The efficiency of harmonic oscillator and spin 1/2 thermal engines is analyzed by decomposing the dynamics into four steps where in two steps the frequency is changing and in the other two steps the system is coupled alternatively to hot and cold reservoirs. The role of the time duration of these steps for getting maximal efficiency was analyzed. Since the for steps for the thermal engines describe non-cyclic circuits only dynamical phases were obtained. Geometric phases can be obtained under special conditions: a) In stochastic chemical kinetics for treating the coupling of a quantum dot with two reservoirs, we have many modulating parameters, and the dynamic is not decomposed into the above steps so that geometric phases can be obtained. b) For getting geometric phases in non-cyclic circuits one should assume that at least two parameters are changed at the same time and that the analysis should be based on a special method which is gauge invariant.